\documentclass[pre,twocolumn,showpacs,superscriptaddress]{revtex4}

\usepackage{graphicx}
\usepackage{amsmath}
\usepackage{amssymb}
\usepackage{mathrsfs}
\usepackage[british]{babel}

\begin{document}

\title{The RARE model: a generalized approach to random relaxation
processes in disordered systems}

\author{Iddo Eliazar}
\affiliation{Holon Institute of Technology, P.O. Box 305, Holon 58102, Israel}
\email{eliazar@post.tau.ac.il}
\author{Ralf Metzler}
\affiliation{Institute for Physics \& Astronomy, University of Potsdam,
14476 Potsdam-Golm, Germany}
\affiliation{Physics Department, Tampere University of Technology,
FI-33101 Tampere, Finland}
\email{rmetzler@uni-potsdam.de}

\begin{abstract}
This paper introduces and analyses a general statistical model, termed
the RARE model, of random relaxation processes in disordered systems. The
model considers excitations, that are randomly scattered around a reaction
center in a general embedding space. The model's input quantities are the
spatial scattering statistics of the excitations around the reaction center,
and the chemical reaction rates between the excitations and the reaction
center as a function of their mutual distance. The framework of the RARE
model is robust, and a detailed stochastic analysis of the random relaxation
processes is established. Analytic results regarding the duration and the
range of the random relaxation processes, as well as the model's
thermodynamic limit, are obtained in closed form. In particular, the case
of power-law inputs, which turn out to yield stretched exponential relaxation
patterns and asymptotically Paretian relaxation ranges, is addressed in detail. 
\end{abstract}


\maketitle

\section{\label{1}Introduction}

Relaxation, the return of a perturbed system into equilibrium, is one of the
most fundamental processes of physical systems. The simplest relaxation model
is the exponential or Debye law $\exp(-t/\tau)$ incorporating the relaxation
time scale $\tau$. In many cases, however, significant deviations from the
exponential law have been observed. The two most prominent generalized
relaxation laws to accommodate these deviations are the stretched exponential
law $\exp(-[t/\tau]^{\alpha})$ with $0<\alpha<1$ and the asymptotic power or
Pareto law (sometimes also called Nutting law) $(1+[t/\tau]^{\beta})^{(-1)}$ with
$\beta>0$. When compared to the simple exponential law, both these generalized
laws correspond to a broad distribution of relaxation times \cite{plonka}.
Generalized relaxation patterns are observed on many scales, ranging from
single molecules \cite{xie,gloeno,volk} over dielectric response \cite{loidl}
to macroscopic viscoelasticity \cite{reiner,schick,gloeno1,rama}.

Theoretical approaches to describe generalized relaxation processes include
parallel relaxation channels \cite{foerster,bluklazu}, hierarchically
constrained dynamics giving rise to complex serial relaxation \cite{palmer},
and defect diffusion models \cite{shlemon}. For the stretched exponential law
Klafter and Shlesinger demonstrated the common universal features behind
these approaches \cite{klashle}. There exist also extensions of stretched
exponentials in models of dynamic relaxation channels \cite{vlad}. Yet
another approach to generalized relaxation dynamics is that of
fractional-order, viscoelastic mechanical bodies
\cite{mainardi,glono,heymans,schiessel,schiessel1}; for their historical
development see Ref.~\cite{mainardi1}. Similar methods based on generalized
dynamic equations and stochastic approaches based on stable distributions
have been used to describe dielectric relaxation behavior
\cite{hilfer,werona}.

In this paper we establish a robust Poissonian approach to complex relaxation
processes based on individual reactions between excitations, that are randomly
scattered around a reaction center in a general embedding space. The resulting
model is termed RARE, the acronym standing for RAndom RElaxations. The
model's input quantities are the spatial scattering statistics of the
excitations around the reaction center, and the chemical reaction rates
between the excitations and the reaction center as a function of their mutual
distance. The RARE model is considerably general, it has a robust framework,
and it allows for a fairly intuitive interpretation of complex relaxation
processes. The RARE model is analyzed in detail, and analytic results
regarding the duration and the range of the model's random relaxation
processes are obtained in closed form.

The paper is organized as follows. The RARE model is
introduced, intuitively explained, and rigorously constructed in section II.
The detailed stochastic analysis of the RARE model and a Monte-Carlo
algorithm for the simulation of the model's random relaxation processes are
presented, respectively, in sections III and IV. The results presented in
section III establish a comprehensive statistical picture of the duration and
the range of the model's random relaxation processes: marginal distributions,
joint distribution, and conditional distributions. The case of power-law inputs,
which are shown to yield stretched exponential relaxation patterns and
asymptotically Paretian relaxation ranges, is addressed in section V. The
thermodynamic limit of the RARE model is investigated in section VI. Detailed
proofs of the results stated along the paper are given in the Appendices.

\section{\label{2}The RARE model}

The RARE model we introduce and explore in this paper is described as
follows. A reaction center is placed at an arbitrary point $P$ of a general
metric space $\mathcal{M}$, and a countable collection of excitations is
scattered randomly across the space $\mathcal{M}$. The excitations
are labeled with the index $i$, and the position of excitation $i$ is the
random point $P_{i}$. The distance in the metric space $\mathcal{M}$ is
measured by a general metric function $\mathbf{d}(\cdot,\cdot)$, and the
distance between the reaction center and excitation $i$ is $D_{i}=\mathbf{d}
(P_{i},P)$. Excitation $i$ is equipped with a random timer $T_{i}$, and a
reaction between the center and the excitations
occurs upon the first timer-expiration event. The
model is statistically characterized by two random variables, the \emph{
reaction time\/} $T$ and the \emph{reaction range\/} $X$, which are defined
as follows: The \emph{reaction time\/} $T$ is the time elapsing until the
first timer expires:
\begin{equation}
\label{201}
T=\min_{i}\{T_{i}\}.
\end{equation}
The \emph{reaction range\/} $X$ is the distance between the reaction center
and the excitation whose timer first expired:
\begin{equation}
\label{202}
X=\sum_{i}D_{i}\mathbf{I}(T=T_{i}).
\end{equation}
In Eq.~\ref{202} and hereinafter, $I(E)$ denotes the indicator function of an
event $E$ (i.e., $I(E)=1$ if the event did occur, and $I(E)=0$ if the event did
not occur).

Setting the space $\mathcal{M}$ in which the reaction takes place to be a
metric space yields great versatility. Indeed, $\mathcal{M}$ can be an
Euclidean space of arbitrary dimension, a non-Euclidean space such as an
elliptic space or a hyperbolic space, a general surface or landscape, a
fractal object, a network, etc. We now turn to specify the RARE model
assumptions, and thereafter present a preliminary analysis.

\subsection{Model assumptions}

To quantify the RARE model the distributions of the random points $\{P_{i}\}$,
as well as the distribution of the random timers $\{T_{i}\}$, need to be
specified. A highly applicable and effective statistical methodology to model
the random scattering of points in general spaces are \emph{Poisson processes\/}
\cite{Kin}. Poisson processes have a wide span of applications ranging from
insurance and finance \cite{EKM} to queuing systems \cite{Wol}, and from
fractals \cite{LT} to power-laws \cite{PWPL}. We henceforth assume that the
random points $\{P_{i}\}$ form a general Poisson process defined on the metric
space $\mathcal{M}$. Consequently, the `displacement theorem' of the theory of
Poisson processes (see section 5.5 in Ref.~\cite{Kin}) implies that the
distances $\{D_{i}\}$ form a general Poisson process defined on the positive
half-line $(0,\infty)$. In what follows we denote by $\rho(x)$ the average
number of excitations, which are within a distance $x$ of the reaction center,
\begin{equation}
\label{203}
\rho(x)=\mathbf{E}\left[\sum_{i}\mathbf{I}(D_{i}\leq x)\right],\,\,\,
x\geq 0.
\end{equation}
In Eq.~(\ref{203}) and in the following, $\mathbf{E}$ denotes the mathematical
expectation. Namely, if $\xi $ is a real-valued random variable
governed by the probability density function $f_{\xi}(x)$ ($x$ real), and
$\phi(x)$ is a real-valued function defined on the real line, then $\mathbf{
E}[\phi(\xi)]=\int_{-\infty }^{\infty }\phi(x)f_{\xi}(x)dx$. Moreover, in
Eq.~(\ref{203})
the function $\rho(x)$ is monotone increasing, and is assumed to start at
zero: $\rho(0)=0$. We note that the derivative $\rho'(x)$ is the
\emph{Poissonian intensity\/} of the Poisson process $\{D_{i}\}$ \cite{Kin}.

As noted in the introduction,
the most common statistical law for reactions in the physical sciences is
the exponential law. Given the distances $\{D_{i}\}$ we now assume that the
timers $\{T_{i}\}$ are independent and exponentially distributed random
variables, and that the exponential distribution of timer $T_{i}$ is
determined by the distance $D_{i}$. In what follows we denote by $\eta(x)$
the exponential rate of the timers as a function of the distance variable $x$.
Namely, given the distance $D_{i}$, the timer $T_{i}$ is exponentially
distributed with tail distribution function
\begin{equation}
\label{204}
\Pr(T_{i}>t|D_{i})=\exp\{-\eta(D_{i})t\}, \,\,\, t\geq 0.
\end{equation}
In other words, given the distance $D_{i}$, the timer $T_{i}$ is exponentially
distributed with mean $\mathbf{E}[T_{i}|D_{i}]=1/\eta(D_{i})$. Typically,
the function $\eta(x)$ is monotone decreasing in the distance variable $x$.

The RARE model is quantified by the pair of functions introduced above, the
\emph{scattering function\/} $\rho(x)$ and the \emph{reactivity function\/}
$\eta(x)$. The scattering function $\rho(x)$ quantifies the underlying spatial
scattering of the excitations, and the reactivity function $\eta(x)$ quantifies
the underlying distance-dependent reaction rate. The `inputs' of the RARE model
are the scattering function $\rho(x)$ and the reactivity function $\eta(x)$,
and the `outputs' of the RARE model are the reaction time $T$ and the reaction
range $X$. In what follows we establish the statistics of the random
outputs based on the given deterministic inputs.

The RARE model introduced herein can be viewed as a descendant of a unified
donor-acceptor energy transfer model presented by Blumen \cite{Blu}. In its
basic form this model considers an $N$-site lattice, in which each site is
occupied by an excitation with probability $p$ (independent of all other
sites), and the reactions between the excitations and the reaction center are
governed by Eq.~(\ref{204}). Thomas and coworkers \cite{thomas} consider a
model similar
to the one developed herein, but with a uniform distribution of excitations.
The RARE model generalizes these models,
as it (i) implements the notion of Poisson processes, (ii) regards both the
reaction time $T$ and the reaction range $X$, and (iii) considers a non-uniform
distribution of excitations. On the one hand, the
implementation of a Poisson scattering of excitations in a general metric space
allows for a high variability and robustness; on the other hand, this Poisson
modeling is highly tractable, as it quantifies all the spatial and scattering
details (both potentially having infinitely many degrees of freedom) into one
single function, the scattering function $\rho(x)$. Moreover, in this paper
we provide a stochastic analysis of the reaction pair $(T,X)$, whereas in
Ref.~\cite{Blu} only the reaction time $T$ was analyzed.

\subsection{Preliminary analysis}

In the next section we present a detailed stochastic analysis of the RARE
model. To facilitate the stochastic analysis of section \ref{3} we present
here a preliminary analysis. In what follows we set $R(x)$ to denote the
aggregate exponential rate corresponding to excitations, whose distance
from the reaction center is greater than $x$,
\begin{equation}
\label{1050}
R(x)=\sum_{i}\eta(D_{i})\mathbf{I}(D_{i}>x),\,\,\, x\geq 0.
\end{equation}
Note that the aggregate rate $R(x)$ is a stochastic process parameterized by
the distance variable $x$.

Conditioned on the realizations of distances $\{D_{i}\}$, the aggregate rate
$R(x)$ yields compact formulas for the conditional distributions of the
reaction time $T$ and the reaction range $X$. Indeed, the assumptions of the
RARE model, combined with the statistical properties of the minima of
independent exponential random variables \cite{Eli}, implies that: (i)
The conditional distribution of the reaction time $T$ is given by the tail
distribution function
\begin{equation}
\label{1051}
\Pr(T>t|\{D_{i}\})=\exp\{-tR(0)\}, \,\,\, t\geq 0.
\end{equation}
(ii) The conditional distribution of the reaction range $X$ is given by the
tail distribution function 
\begin{equation}
\label{1052}
\Pr(X>x|\{D_{i}\})=\frac{R(x)}{R(0)},\,\,\, x\geq 0.
\end{equation}
(iii) The joint conditional distribution of the reaction pair $(T,X)$ is given
by the joint tail distribution function 
\begin{equation}
\label{1053}
\Pr(T>t,X>x|\{D_{i}\})=\exp\{-tR(0)\}\frac{R(x)}{R(0)},\,\,\, t,x\geq 0.
\end{equation}

Note that given the realizations of the distances $\{D_{i}\}$ the reaction time
$T$
and the reaction range $X$ turn out to be \emph{independent\/} random variables.
Indeed, the joint tail distribution function of the reaction pair $(T,X)$ equals
the product of the tail distribution functions of the reaction time $T$ and the
reaction range $X$,
\begin{eqnarray}
\nonumber
\Pr(T>t,X>x|\{ D_{i}\})&=&\\
&&\hspace*{-3.6cm}\Pr(T>t|\{ D_{i}\})\Pr(X>x|\{ D_{i}\}),\,\,\, t,x\geq 0.
\label{205}
\end{eqnarray}

\section{Stochastic analysis}
\label{3}

The preliminary analysis presented in section \ref{2} provides us with the
conditional distribution of the reaction pair $(T,X)$, conditioned on the
realizations of distances $\{D_{i}\}$. A stochastic analysis detailed the
Appendices A-D shifts us from the aforementioned conditional distribution to the
(unconditional) distribution of the reaction pair $(T,X)$. In this section
we present the key results of this stochastic analysis.

\textbf{Reaction time}. The tail distribution function of the reaction time
$T$ is given by
\begin{equation}
\label{411}
\Pr(T>t)=\exp\left(-\int_{0}^{\infty }[1-\exp\{-\eta(x)t\}]\rho(dx)\right),
\end{equation}
for $t\geq 0$. In turn, taking the limit $t\rightarrow\infty$ in Eq.~(\ref{411})
yields the probability that a reaction never takes place,
\begin{equation}
\label{410}
\Pr(T=\infty)=\exp\{-\rho(\infty)\}.
\end{equation}%
Eq.~(\ref{410}) implies that a reaction takes place with certainty, i.e.,
$\Pr(T<\infty)=1$, if and only if the scattering function $\rho(x)$ diverges,
which in turn occurs if and only if there are infinitely many excitations.
Moreover, differentiating Eq.~(\ref{411}) with respect to the time variable
$t$ yields the probability density function of the reaction time $T$,
\begin{equation}
\label{412}
f_{T}(t)=\Pr(T>t)\int_{0}^{\infty}\exp\{-\eta(x)t\}\eta(x)\rho(dx),\,\,\,
t>0.
\end{equation}
It is straightforward to deduce from Eq.~(\ref{412}) that the probability
density function $f_{T}(t)$ is monotone decreasing from the level $f_{T}(0)=
\int_{0}^{\infty }\eta(x)\rho(dx)$ (which can be either finite or infinite)
to the level $f_{T}(\infty)=0$.

\textbf{Temporal hazard rate}. The \emph{hazard rate\/} of reaction time $T$
is defined in terms of
\begin{equation}
\label{321}
h_{T}(t)=\lim_{\delta\rightarrow0}\frac{1}{\delta}\Pr(T\leq t+\delta|T>t)=
\frac{f_{T}(t)}{\Pr(T>t)},
\end{equation}
with $t>0$. Namely, the hazard rate $h_{T}(t)$ is the realization rate of the
random variable $T$ at time $t$, provided that it has not already occurred up
to time $t$. The hazard rate is most commonly used in applied probability and
in reliability theory \cite{Ros1,Tijms,BP}. Substituting the tail distribution
function $\Pr(T>t)$ of Eq.~(\ref{411}) and the probability density function
$f_{T}(t)$ of Eq.~(\ref{412}) into Eq.~(\ref{321}) yields the hazard rate of
the reaction time $T$,
\begin{equation}
\label{322}
h_{T}(\tau)=\int_{0}^{\infty }\exp\{-\eta(x)t\}\eta(x)\rho(dx),\,\,\,
t>0.
\end{equation}
It is evident from Eq.~(\ref{322}) that the hazard rate $h_{T}(t)$, analogous
to the probability density function $f_{T}(t)$, is monotone decreasing from
the level $h_{T}(0)=\int_{0}^{\infty}\eta(x)\rho(dx)$ (which can be either
finite or infinite) to the level $h_{T}(\infty)=0$. Note that a monotone
decreasing hazard rate $h_{T}(t)$ implies that the realization rate of the
reaction time $T$ \emph{diminishes\/} in the course of time. Namely, the
longer we are waiting for the reaction time $T$ to realize, the \emph{lower\/}
the likelihood that it occurs immediately.

\textbf{Reaction range}. The probability density function of the reaction
range $X$ is given by
\begin{equation}
\label{421}
f_{X}(x)=(1-\mathbf{E}[\exp\{-\eta(x)T\}])\rho'(x),\,\,\, x>0.
\end{equation}
Eq.~(\ref{421}) implies that if the Poissonian intensity $\rho'(x)$ is a
monotone decreasing function then so is the probability density function
$f_{X}(x)$. In general, however, the probability density function $f_{X}(x)$
itself is not necessarily monotone decreasing. Eq.~(\ref{421}) further
implies that the tail distribution function of the reaction range $X$ is
given by
\begin{equation}
\label{422}
\Pr(X>x)=\int_{x}^{\infty}(1-\mathbf{E}[\exp\{-\eta(y)T\}])\rho(dy),\,\,\,
x\geq 0.
\end{equation}
Note that both the probability density function $f_{X}(x)$ and the tail
distribution function $\Pr(X>x)$ of the reaction range $X$ involve the
Laplace transform of the reaction time $T$.

\textbf{Reaction pair}. The joint probability density function of the
reaction pair $(T,X)$ is given by
\begin{equation}
\label{611}
f_{(T,X)}(t,x)=\Pr(T>t)\exp\{-t\eta(x)\}\eta(x)\rho'(x),\,\,\, t,x>0.
\end{equation}
Comparing the joint probability density function $f_{(T,X)}(t,x)$ from
Eq.~(\ref{611}) with the product $f_{T}(t)f_{X}(x)$ of the probability
density function $f_{T}(t)$ from Eq.~(\ref{412}) and the probability density
function $f_{X}(x)$ from Eq.~(\ref{421}), it is straightforward to see that
\begin{equation}
\label{612}
f_{(T,X)}(t,x)\neq f_{T}(t)f_{X}(x),\,\,\, t,x>0.
\end{equation}%
Relation (\ref{612}) implies that the reaction time $T$ and the reaction range
$X$ are \emph{dependent\/} random variables. This dependence is diametric to
the \emph{conditional independence\/} of the reaction time $T$ and the reaction
range $X$, conditioned on the realizations of distances $\{ D_{i}\}$, which is
manifested by Eq.~(\ref{205}). Thus the random Poissonian structure of the
distances $\{D_{i}\}$ induces a statistical dependence between the reaction time
$T$ and the reaction range $X$.

\textbf{Conditional distributions}. The conditional distribution of the
reaction time $T$, conditioned on the realization of the reaction range $X$,
is given by the tail distribution function
\begin{equation}
\label{415}
\Pr(T>t|X=x)=\frac{\int_{t}^{\infty}\Pr(T>s)\exp\{-\eta(x)s\}ds}{\int_{0}^{
\infty}\Pr(T>s)\exp\{-\eta(x)s\}ds},
\end{equation}
for $t>0$. Also, the conditional distribution of the reaction range $X$,
conditioned to the realization of the reaction time $T$, is given by the
tail distribution function
\begin{equation}
\label{425}
\Pr(X>x|T=t)=\frac{\int_{x}^{\infty}\exp\{-\eta(y)t\}\eta(y)\rho(dy)}{\int_{0}^{
infty}\exp\{-\eta(y)t\}\eta(y)\rho(dy)},
\end{equation}
for $x>0$.

As noted at the end of section II.A, the RARE model can be viewed as a
descendant of Blumen's unified donor-acceptor energy transfer model
\cite{Blu} and Thomas et al.'s donor-acceptor recombination model \cite{thomas}.
In Refs.~\cite{Blu,thomas} the reaction time $T$ was analyzed, and a
counterpart of Eq.~(\ref{411}) was established (in the context of Bernoulli
scattering of excitations over an $N$-site lattice). The analysis carried
out in this section goes significantly beyond those of Refs.~\cite{Blu,thomas}
as it
addresses the marginal, joint, and conditional distributions of the reaction
pair $(T,X)$. Moreover, all results established in this section are valid in
the context of general Poisson scattering of excitations over general metric
spaces, a setting which allows for high variability and robustness, whilst
yielding closed-form analytic expressions.

\section{Monte-Carlo simulation}
\label{4}

The stochastic analysis presented in the previous section facilitates the
construction of a numerical Monte-Carlo algorithm for the simulation of the
reaction pair $(T,X)$. The steps of the Monte-Carlo algorithm are as follows:

(1) Identify the inputs of the RARE model: the scattering function $\rho(x)$
and the reactivity function $\eta(x)$.

(2) Using the scattering function $\rho(x)$ and the reactivity function $\eta
(x)$, numerically compute the integral
\begin{equation}
\label{621}
I(t,x)=\int_{0}^{x}\exp\{-\eta(y)t\}\eta(y)\rho(dy),\,\,\, t,x\geq 0.
\end{equation}

(3) Using the integral $I(t,x)$, numerically compute the cumulative
distribution function of the reaction time
\begin{equation}
\label{622}
F(t)=1-\exp\left(-\int_{0}^{t}I(\tau,\infty)d\tau\right),\,\,\, t\geq 0,
\end{equation}%
and then numerically compute its inverse function $F^{-1}(u)$ ($0\leq u\leq1$).

(4) Using the integral $I(t,x)$, numerically compute the conditional
cumulative distribution function of the reaction range
\begin{equation}
\label{623}
G(x;t)=\frac{I(t,x)}{I(t,\infty)},\,\,\, x\geq 0,
\end{equation}
and then numerically compute its inverse function $G^{-1}(u;t)$ ($0\leq u\leq
1$).

(5) Generate a pair $(U_{1},U_{2})$ of independent random variables, which
are uniformly distributed over the unit interval, and generate the reaction
pair $(T,X)$ via
\begin{equation}
\label{624}
T=F^{-1}(U_{1})\mbox{ and }X=G^{-1}(U_{2};F^{-1}(U_{1})).
\end{equation}

The third step of the Monte-Carlo algorithm is based on Eq.~(\ref{411}), the
fourth step is based on Eq.~(\ref{425}), and in the fifth step we applied
the following basic simulation principle \cite{Ros2}: If $\Phi(x)$ ($x\geq 0$) is
the cumulative distribution function of a positive-valued random variable $\xi$,
and if $U$ is a random variable which is uniformly distributed over the unit
interval (and independent of $\xi$), then the random variable $\Phi^{-1}(U)$ is
equal in law to the random variable $\xi$.

The Monte-Carlo algorithm can be further used to simulate \emph{random walks\/}
whose dynamics are governed by the RARE model. To that end assume that the
scattering of the excitations is spatially homogeneous. This homogeneity
implies that the Poissonian structure of the distances $\{D_i\}$ is
invariant with respect to the position $P$ of the reaction center. Now consider
the following reaction-based propagation scheme: if the reaction center reacted
with excitation $i$ then the center jumps from its initial position $P$ to the
position of excitation $i$, $P_i$, and thereafter the process starts anew.
This propagation scheme generates a renewal \emph{Continuous Time Random Walk\/}
(CTRW) \cite{MoW,MK,KS}, whose law of motion is as follows:

(1) Initiate from an arbitrary position $P_0$ at time $t_0$, and
simulate a reaction pair $(T,X)$.

(2) At time $t_1=t_0+T$ move to a point $P_1$ which is uniformly distributed
on a sphere with radius $X$ centered at the point $P_0$.

(3) Set $t_0:=t_1$ and $P_0:=P_1$, and go back to step (1).

The reaction time $T$ is the CTRW's generic waiting time, and the reaction
range $X$ is CTRW's generic jump length. We emphasize that since the reaction
time $T$ and the reaction range $X$ are \emph{dependent\/} random variables,
the RARE model induces CTRWs with \emph{coupled\/} waiting times and jump
lengths \cite{SKW1,SKW2,SKW3}.

\section{Power-law inputs}
\label{5}

To illustrate the results established so far we consider now the example of
a RARE process with power-law inputs. Specifically, we consider both the
scattering function and the reactivity function to be power-laws
\begin{equation}
\label{500}
\rho(x)=ax^{\alpha}\mbox{ and }\eta(x)=bx^{-\beta},\,\,\, x>0
\end{equation}
whose coefficients $a$ and $b$ as well as the exponents $\alpha$ and $\beta$
are all positive parameters.

Power-law inputs occur naturally in many cases. If we scatter the excitations
uniformly across $d$-dimensional Euclidean space using a spatially homogeneous
Poisson process, then the scattering function $\rho(x)$ will be a power-law
with exponent $\alpha=d$. Indeed, if the excitations are scattered uniformly
across the $d$-dimensional Euclidean space, then the mean number of excitations 
present in a ball of radius $x$, quantified by the scattering function $\rho(x)$,
is proportional to $x^d$. Moreover, in many fractal settings, which are
abundant in the physical sciences, the scattering of the chemical agents is
uniform across some fractal unbounded subset of the $d$-dimensional Euclidean
space. In such fractal settings the scattering function $\rho(x)$ will be a
power-law with exponent $\alpha$, that equals the fractal dimension of the
underlying fractal subset \cite{falconer}. On the other hand, power-law decays
of physical and chemical interactions as a function of the distance between the
interacting elements are prevalent in the the physical sciences, the best known
examples being Newton's law of gravitation, Coulomb's law, or the van der
Waals law.

We turn now to describe the statistical behavior of the RARE model with
power-law inputs. For the RARE model to be well-defined, the exponent $\alpha$
of the scattering function must be smaller than the exponent $\beta$ of the
reactivity function, and thus we assume that $\alpha<\beta $.

\textbf{Reaction time}. From Eq.~(\ref{411}) we obtain that the tail
distribution function of the reaction time $T$ is given by
\begin{equation}
\label{501}
\Pr(T>t)=\exp\left(-c_{1}t^{\alpha/\beta}\right),\,\,\, t\geq0,
\end{equation}
where the precise value of the coefficient $c_{1}$ is given by $c_{1}=\Gamma(
1-\alpha/\beta)ab^{\alpha/\beta}$. The tail distribution function of
Eq.~(\ref{501}) characterizes a \emph{stretched exponential\/} law
\cite{Koh,WW,KC}. The hazard rate of the reaction time $T$ is a power-law with
a negative exponent equal to $-(1-\alpha/\beta)$, and the moments of the
reaction time $T$ are given by
\begin{equation}
\label{502}
\mathbf{E}[T^{m}]=\Gamma(1+m\beta/\alpha)c_{1}^{-m\beta/\alpha},\,\,\, m>0.
\end{equation}
We emphasize that although the reaction time $T$ has convergent moments of all
orders, its moment generating function is divergent: $\mathbf{E}[\exp(\theta T
)]=\infty$ for all $\theta>0$. This statistical behavior of the stretched
exponential distribution, convergent moments and divergent moment generating
function, implies that the reaction time $T$ displays a form of randomness
which Mandelbrot categorized as `borderline randomness' \cite{Man}. Another
well-known probability distribution displaying such borderline randomness is
the log-normal distribution \cite{Mit}.

\textbf{Reaction range}. Evaluating Eq.~(\ref{422}), while applying a
moment-expansion of the Laplace transform of the reaction time $T$, we
obtain that the tail distribution function of the reaction range $X$ is
given by the power-expansion
\begin{equation}
\label{503}
\Pr(X>x)=a\alpha\sum_{m=1}^{\infty}(-1)^{m+1}\frac{b^m\mathbf{E}[T^m]}{m!
(m\beta-\alpha)}\frac{1}{x^{m\beta-\alpha}},
\end{equation}
with $x>0$. In turn, Eq.~(\ref{503}) implies that the asymptotic behavior of
the tail distribution function of the reaction range $X$ is given by 
\begin{equation}
\label{504}
\Pr(X>x)\approx\frac{c_2}{x^{\beta-\alpha}},\,\,\, x\rightarrow\infty,
\end{equation}%
where the coefficient $c_2$ is given by $c_2=[\Gamma(1+\beta/\alpha)a^{1-\beta
/\alpha}]/[\Gamma(1-\alpha/\beta)(\beta-\alpha)]$. The tail asymptotics of
Eq.~(\ref{504}) characterize an \emph{asymptotically Paretian\/} distribution
with exponent $\beta-\alpha$ \cite{PWPL,Par,New,CSN}. The moments of the
reaction range $X$ are convergent, i.e., $\mathbf{E}[X^m]<\infty$, if and only
if the exponent $m$ is in the range $0<m<\beta-\alpha$. In particular, the
reaction range $X$ has a convergent mean if and only if $1+\alpha<\beta$. This
statistical behavior of asymptotically Paretian distributions, convergent
moments only up to a given order, implies that the the reaction range $X$
displays a form of randomness which Mandelbrot categorized as `wild
randomness' \cite{Man,MT}. For a recent treatment of the `categorization of
randomness' see Ref.~\cite{elico}.

\textbf{Conditional distribution}. From Eq.~(\ref{425}) we obtain that the
conditional distribution of the reaction range $X$, conditioned on the
realization of the reaction time $T$, is given by the tail distribution
function
\begin{equation}
\label{505}
\Pr(X>x|T=t)=\int_0^{tbx^{-\beta }}\frac{\exp(-u)u^{-\alpha/\beta}}{\Gamma(
1-\alpha/\beta)}du,\,\,\, x>0.
\end{equation}%
Note that the integrand on the right hand side of Eq.~(\ref{505}) is the
probability density function of a \emph{Gamma distribution\/} with exponent
$1-\alpha/\beta $. Thus, if we set $\xi$ to be a Gamma-distributed random
variable with exponent $1-\alpha/\beta$, we obtain that $\Pr(X>x|T=t)=\Pr
(\xi\leq tbx^{-\beta})$. Eq.~(\ref{505}) further implies that the asymptotic
behavior of the conditional tail distribution function of the reaction range
$X$ is given by
\begin{equation}
\label{506}
\Pr(X>x|T=t)\approx c_3\frac{t^{1-\alpha/\beta}}{x^{\beta-\alpha}},\,\,\,
x\rightarrow\infty,
\end{equation}%
where $c_3=b^{1-\alpha/\beta}/\Gamma(2-\alpha/\beta)$. As in the case of
Eq.~(\ref{504}), the tail asymptotics of Eq.~(\ref{506}) characterize an
asymptotically Paretian distribution with exponent $\beta-\alpha$.
Consequently, also with any given realization of the reaction time $T$, the
reaction range $X$ displays wild randomness.

\section{Thermodynamic limit}
\label{6}

In this section we explore the statistical behavior of the RARE model as the
\emph{concentration\/} of the excitations is increased to infinity.
Specifically, we increase by $n$-fold the concentration of the excitations,
and examine the limiting statistical behavior of the reaction time and the
reaction range in the limit $n\rightarrow\infty$.

An $n$-fold increase of the concentration of the excitations results in
replacing the `original' scattering function $\rho(x)$ by the `concentrated'
scattering function $\rho_n(x)=n\rho(x)$. On the other hand, an $n$-fold
increase of the concentration of the excitations does not affect the
reactivity function $\eta(x)$. In this section we denote by $(T_n,X_n)$ the
reaction pair corresponding to a RARE model with scattering function $\rho_n
(x)$ and reactivity function $\eta(x)$. Moreover, we set 
\begin{equation}
\label{600}
\lambda=\int_{0}^{\infty}\eta(x)\rho(dx),
\end{equation}
and assume that the integral on the right-hand-side of Eq.~(\ref{600}) is
convergent.

We introduce the scaled reaction time $S_n=nT_n$. A stochastic limit analysis
detailed in Appendix E asserts that the joint probability density function
of the reaction pair $(S_n,X_n)$ attains the limit
\begin{equation}
\label{601}
\lim_{n\rightarrow\infty}f_{(S_n,X_n)}(s,x)=\exp(-\lambda s)\eta(x)\rho'(x),
\,\,\, s,x>0.
\end{equation}
Namely, the reaction pair $(S_n,X_n)$ converges in law (as $n\rightarrow\infty$)
to a limiting random vector $(S_{\infty},X_{\infty})$, whose distribution is
governed by the joint probability density function
\begin{equation}
\label{602}
f_{(S_{\infty},X_{\infty})}(s,x)=\lambda\exp(-\lambda s)\frac{1}{\lambda}\eta
(x)\rho'(x),\,\,\, s,x>0.
\end{equation}%
Eq.~(\ref{602}) implies the following:

(1) The scaled reaction time $S_n$ converges in law to a stochastic limit
$S_{\infty}$, which is \emph{exponentially distributed\/} with rate $\lambda$
and density function 
\begin{equation}
\label{603}
f_{S_{\infty}}(s)=\lambda\exp(-\lambda s),\,\,\, s>0.
\end{equation}

(2) The reaction range $X_n$ converges in law to a stochastic limit $X_{\infty}$
whose distribution is governed by the probability density function
\begin{equation}
\label{604}
f_{X_{\infty}}(x)=\frac{1}{\lambda}\eta(x)\rho'(x),\,\,\, x>0.
\end{equation}

(3) The joint probability density function of the random vector $(S_{\infty },
X_{\infty})$ equals the product of its corresponding marginal probability
density functions
\begin{equation}
\label{605}
f_{(S_{\infty},X_{\infty})}(s,x)=f_{S_{\infty}}(s)f_{X_{\infty}}(x),\,\,\,
s,x>0,
\end{equation}
and hence the stochastic limits $S_{\infty}$ and $X_{\infty}$ are
\emph{independent\/} random variables.

The stochastic limit result established in this section gets us `all around
the circle'. Indeed, in the preliminary analysis of section \ref{2} we obtained
that the conditional distribution of the reaction time $T$, conditioned
on the realizations of distances $\{D_i\}$, is exponential and independent of
the reaction range $X$. However, in section \ref{3} we found that the
distribution of the reaction time $T$ is general, and is tightly coupled to
the distribution of the reaction range $X$. Thus, the random Poissonian
structure of the distances $\{D_i\}$ shifts the distribution of the reaction
time $T$ from exponential to general, and induces a dependence between the
reaction time $T$ and the reaction range $X$. In this section we
established that in the infinite concentration limit $n\rightarrow\infty$
the original statistical structure is recovered: (i) the exponential
distribution is restored, as the stochastic limit $S_{\infty}$ of the scaled
reaction time is exponentially distributed; and (ii) the independence is
restored, as the stochastic limit $S_{\infty}$ of the scaled reaction time
and the stochastic limit $X_{\infty}$ of the reaction range are independent
random variables.

To illustrate the stochastic limit result established in this section consider
the example of power law scattering functions studied in the previous section
and exponential reactivity functions
\begin{equation}
\label{606}
\rho(x)=ax^{\alpha}\mbox{ and }\eta(x)=b\exp(-\beta x),\,\,\, x>0,
\end{equation}
where the coefficients $a$ and $b$ as well as the parameters $\alpha$ and
$\beta$, are all positive. In this example the stochastic limit $S_{\infty}$
is exponentially distributed with rate 
\begin{equation}
\label{608}
\lambda=ab\frac{\Gamma(1+\alpha)}{\beta^{\alpha}},
\end{equation}
and the stochastic limit $X_{\infty}$ is \emph{Gamma distributed\/} with
probability density function
\begin{equation}
\label{607}
f_{X_{\infty}}(x)=\frac{\beta^{\alpha}}{\Gamma(\alpha)}\exp(-\beta x)x^{\alpha
-1},\,\,\, x>0.
\end{equation}
Note that the Gamma probability density function of Eq.~(\ref{607}) is (i)
unbounded and monotone decreasing in the exponent range $\alpha <1$; (ii)
bounded and monotone decreasing at the exponent value $\alpha=1$, in
which case $X_{\infty}$ is exponentially distributed with mean $1/\beta $;
(iii) bounded and unimodal, with mode $x_{\ast}=(\alpha-1)/\beta $, in the
exponent range $\alpha >1$.

\section{Conclusion}
\label{7}

In this paper we presented a general spatio-chemical stochastic model for
generalized RAndom Relaxation (RARE) processes in complex disordered systems.
The RARE model considers a collection of excitations which are randomly
scattered around a reaction center in some general embedding metric space.
The RARE model has two input quantities:  (i) the scattering function
$\rho(x)$ quantifying the scattering intensity of the excitations around the
reaction center as a function of the distance ($x>0$) from the center; (ii) the
reactivity function $\eta(x)$ quantifying the reaction rate between the
excitations and the reaction center as a function of the distance ($x>0$) from
the center. The scattering and reactivity functions provide a straightforward
intuitive description, as well as a precise mathematical formulation, of
general relaxation processes in complex disordered systems.

The RARE model has two random outputs: (i) the reaction time $T$ of its
random relaxation process; (ii) the reaction range $X$ of its random
relaxation process. A detailed stochastic analysis of the reaction pair
$(T,X)$ was carried out, yielding closed form results regarding the
statistics of this pair: marginal distributions, joint distribution, and
conditional distributions. The results established further led to a
Monte-Carlo algorithm for the simulation of the model's random relaxation
process, and to the thermodynamic of the RARE model. In addition, we
investigated in detail the case of power-law inputs, which were shown to
yield stretched exponential relaxation patterns and asymptotically Paretian
relaxation ranges. 

The RARE model is a compact and transparent stochastic approach to
non-exponential
relaxation processes. On the one hand the model's inputs are both intuitively
clear and mathematically precise, and are directly related to the physical
properties of the system considered. On the other hand the model is rather
robust and versatile, as its general mathematical formulation accommodates
diverse physical settings. We therefore expect the RARE model to be useful to
a wide range of applications in the physical sciences and beyond.

\acknowledgments

RM acknowledges funding from the Academy of Finland (FiDiPro scheme).

\begin{appendix}
\label{8}

\section{The distribution of the reaction time~$T$}

Eq.~(\ref{201}) implies that the tail distribution function of the reaction
time $T$ is given by 
\begin{equation}
\label{1021}
\Pr(T>t)=\Pr\left(\min_i\{T_i\}>t\right)=\Pr\left(\bigcap_i\{T_i>t\}\right).
\end{equation}
After conditioning with respect to the distances $\{D_i\}$, this is equal to
\begin{equation}
\label{1022}
\Pr(T>t)=\mathbf{E}\left[\Pr\left(\bigcap_i\{T_i>t\}|\{D_i\}\right)\right].
\end{equation}
Using the assumptions of the RARE model and Eq.~(\ref{204})), we have
\begin{eqnarray}
\nonumber
\Pr(T>t)&=&\mathbf{E}\left[\prod_i\Pr(T_i>t|D_i)\right]\\
&=&\mathbf{E}\left[\prod_i\exp\{-\eta(D_i)t\}\right].
\end{eqnarray}
With Campbell's theorem of the theory of Poisson processes (see Section
3.2 in Ref.~\cite{Kin}),
\begin{equation}
\label{1023}
\Pr(T>t)=\exp\left(-\int_0^{\infty}[1-\exp\{-\eta(x)t\}]\rho(dx)
\right).
\end{equation}
Eqs.~(\ref{1021}) to (\ref{1023}) prove Eq.~(\ref{411}).

\section{The distribution of the reaction pair~$\left( T,X\right) $}

In what follows we set
\begin{equation}
\label{1054}
E(t,\theta,x)=\mathbf{E}[\exp\{-tR(0)-\theta R(x)\}],\,\,\, t,\theta ,x\geq 0.
\end{equation}
Eq.~(\ref{1050}) implies that
\begin{eqnarray}
\nonumber
tR(0)+\theta R(x)&=&\sum_i\Big(t\eta(D_i)\mathbf{I}(D_i\leq x)\\
&&+(t+\theta)\eta(D_i)\mathbf{I}(D_i>x)\Big),
\label{1055}
\end{eqnarray}
and hence Eq.~(\ref{1054}) further implies that
\begin{eqnarray}
\nonumber
E(t,\theta,x)&=&\mathbf{E}\Big[\prod_i\exp\Big(-t\eta(D_i)\mathbf{I}(D_i\leq
x)\\
&&-(t+\theta)\eta(D_i)\mathbf{I}(D_i>x)\Big)\Big] 
\label{1056}
\end{eqnarray}
Consequently, Campbell's theorem of the theory of Poisson processes (see
Section 3.2 in Ref.~\cite{Kin}) implies that
\begin{equation}
\label{1057}
E(t,\theta,x)=\exp\{-\Psi(t,\theta,x)\},
\end{equation}
where
\begin{widetext}
\begin{eqnarray}
\nonumber
\Psi(t,\theta,x)&=&\int_0^{\infty}\Big[1-\exp\{-t\eta(s)\mathbf{I}(s\leq x)
-(t+\theta)\eta(s)\mathbf{I}(s>x)\}\Big]\rho(ds)\\ 
&=&\int_0^{x}\Big[1-\exp\{-t\eta(s)\}\Big]\rho(ds)+\int_x^{\infty}\Big[1-
\exp\{-(t+\theta)\eta(s)\}\Big]\rho(ds).
\label{1058}
\end{eqnarray}
\end{widetext}

Applying conditioning with respect to the distances $\{D_i\}$ and using
Eq.~(\ref{1053}), we obtain that the joint tail distribution function of the
reaction pair $(T,X)$ is given by
\begin{eqnarray}
\nonumber
\Pr(T>t,X>x)&=&\mathbf{E}[\Pr(T>t,X>x|\{D_i\})]\\ 
&&\hspace*{-1.8cm}
=\mathbf{E}\left[\exp\{-tR(0)\}\frac{R(x)}{R(0)}\right],\,\,\, t,x\geq 0.
\label{1061}
\end{eqnarray}
Differentiating this expression with respect to the variable $t$ yields
\begin{equation}
\frac{\partial}{\partial t}\Pr(T>t,X>x)=-\mathbf{E}[\exp\{-tR(0)\}R(x)].
\label{1062}
\end{equation}
Conversely, differentiating Eq.~(\ref{1054}) with respect to the variable
$\theta$ yields
\begin{equation}
\label{1063}
\frac{\partial}{\partial\theta}E(t,\theta,x)=-\mathbf{E}[\exp\{-tR(0)-\theta
R(x)\}R(x)].
\end{equation}
Combination of Eqs.~(\ref{1062}) and (\ref{1063}) yields
\begin{equation}
\label{1064}
\frac{\partial}{\partial t}\Pr(T>t,X>x)=\left.\frac{\partial}{\partial\theta}
E(t,\theta,x)\right|_{\theta=0}.
\end{equation}

Now, Eq.~(\ref{1057}) implies that
\begin{equation}
\label{1065}
\left.\frac{\partial}{\partial\theta}E(t,\theta,x)\right|_{\theta=0}=-\exp\{
-\Psi(t,0,x)\}\left.\frac{\partial}{\partial\theta}\Psi(t,\theta,x)\right|_{
\theta=0},
\end{equation}%
and Eq.~(\ref{1058}) further implies that 
\begin{equation}
\label{1066}
\Psi(t,0,x)=\int_0^{\infty}\Big[1-\exp\{-t\eta(s)\}\Big]\rho(ds),
\end{equation}
and 
\begin{equation}
\label{1067}
\left.\frac{\partial}{\partial\theta}\Psi(t,\theta,x)\right|_{\theta=0}=\int_x
^{\infty}\exp\{-t\eta(s)\}\eta(s)\rho(ds).
\end{equation}
Thus, combining Eqs.~(\ref{1064}) to (\ref{1067}), we conclude that
\begin{widetext}
\begin{equation}
\label{1068}
-\frac{\partial}{\partial t}\Pr(T>t,X>x)=\exp\left(-\int_0^{\infty}\Big[1-\exp
\{-t\eta(s)\}\Big]\rho(ds)\right)\int_x^{\infty}\exp\{-t\eta(s)\}\eta(s)\rho(
ds).
\end{equation}
Finally, differentiating this last expression with respect to the variable $x$
and using Eq.~(\ref{411}), we arrive at the desired result,
\begin{equation}
\label{1069}
\frac{\partial^{2}}{\partial t\partial x}\Pr(T>t,X>x)=\Pr(T>t)\exp\{-t\eta(x)\}
\eta(x)\rho'(x),
\end{equation}
such that Eq.~(\ref{1069}) proves Eq.~(\ref{611}).
\end{widetext}

\section{The marginal distributions of the reaction pair $(T,X)$}

The probability density function of the reaction time $T$ is attained by
integrating the joint probability density function of the random pair
$(T,X)$ over the distance variable $x$,
\begin{equation}
\label{1074}
f_T(t)=\int_0^{\infty}f_{(T,X)}(t,x)dx.
\end{equation}
With Eq.~(\ref{611}),
\begin{equation}
\label{1075}
f_T(t)=\Pr(T>t)\int_0^{\infty}\exp\{-t\eta(x)\}\eta(x)\rho(dx),\,\,\, t>0.
\end{equation}
Eqs.~(\ref{1074}) and (\ref{1075}) prove Eq.~(\ref{412}). Note that the
Laplace transform of the reaction time $T$ is given by
\begin{equation}
\label{1076}
\mathbf{E}[\exp(-\theta T)]=\int_0^{\infty}\exp(-\theta t)f_T(t)dt.
\end{equation}
Integration by parts yields
\begin{equation}
\label{1077}
\mathbf{E}[\exp(-\theta T)]=1-\theta\int_0^{\infty}\exp(-\theta t)\Pr(T>t)dt,
\end{equation}
where $\theta\geq0$.

The probability density function of the reaction range $X$ is attained by
integrating the joint probability density function of the random pair
$(T,X)$ over the time variable $t$,
\begin{equation}
\label{1071}
f_X(x)=\int_0^{\infty}f_{(T,X)}(t,x)dt
\end{equation}
and, using Eq.~(\ref{611}),
\begin{equation}
\label{1072}
f_X(x)=\left(\int_0^{\infty}\Pr(T>t)\exp\{-t\eta(x)\}dt\right)\eta(x)\rho'(x).
\end{equation}
With Eqs.~(\ref{1076}) and (\ref{1077}) and $\theta=\eta(x)$ we find
\begin{equation}
\label{1073}
f_X(x)=(1-\mathbf{E}[\exp\{-\eta(x)T\}])\rho'(x),\,\,\, x>0.
\end{equation}%
Eqs.~(\ref{1071}) to (\ref{1073}) prove Eq.~(\ref{421}).

\section{The conditional distributions of the reaction pair $(T,X)$}

The conditional probability density function of the reaction time $T$,
conditioned to the realization of the reaction range $X$, is given by
\begin{equation}
\label{1085}
f_{T|X=x}(t)=\frac{f_{(T,X)}(t,x)}{f_X(x)}.
\end{equation}
With Eqs.~(\ref{611}) and (\ref{421}),
\begin{eqnarray}
\nonumber
f_{T|X=x}(t)&=&\frac{\Pr(T>t)\exp\{-t\eta(x)\}\eta(x)\rho'(x)}{(\int_0^{\infty}
\Pr(T>\tau)\exp\{-\tau\eta(x)\}d\tau)\eta(x)\rho'(x)}\\
&=&\frac{\Pr(T>t)\exp\{-t\eta(x)\}}{\int_0^{\infty}\Pr(T>\tau)\exp\{-\tau\eta(
x)\}d\tau}.
\label{1086}
\end{eqnarray}
In turn, Eqs.~(\ref{1085}) and (\ref{1086}) imply Eq.~(\ref{415}).

The conditional probability density function of the reaction range $X$,
conditioned to the realization of the reaction time $T$, is given by
\begin{equation}
\label{1087}
f_{X|T=t}(x)=\frac{f_{(T,X)}(t,x)}{f_T(t)}.
\end{equation}
Using Eqs.~(\ref{611}) and (\ref{412}), we see that
\begin{eqnarray}
\nonumber
f_{X|T=t}(x)&=&\frac{\Pr(T>t)\exp\{-t\eta(x)\}\eta(x)\rho'(x)}{\Pr(T>t)\int_0^{
\infty}\exp\{-\eta(x)t\}\eta(x)\rho(dx)}\\
&=&\frac{\exp\{-t\eta(x)\}\eta(x)\rho'(x)}{\int_0^{\infty}\exp\{-\eta(x)t\}
\eta(x)\rho(dx)}.
\label{1088}
\end{eqnarray}
In turn, Eqs.~(\ref{1087}) and (\ref{1088}) imply Eq.~(\ref{425}).

\section{The limit distribution of the reaction pair $(S_n,X_n)$}

The scaling $S_n=nT_n$ implies the following connection between the
joint cumulative distribution function of the reaction pair $(S_n,X_n)$
and the joint cumulative distribution function of the reaction pair
$(T_n,X_n)$,
\begin{equation}
\label{1090}
\Pr(S_n\leq s,X_n\leq x)=\Pr\left(T_n\leq\frac{s}{n},X_n\leq x\right),\,\,\,
s,x>0.
\end{equation}
Consequently, differentiating this expression with respect to the variables $s$
and $x$ yields the following connection between the joint probability density
function of the reaction pair $(S_n,X_n)$ and the joint probability density
function of the reaction pair $(T_n,X_n)$,
\begin{equation}
\label{1091}
f_{(S_n,X_n)}(s,x)=f_{(T_n,X_n)}\left(\frac{s}{n},x\right)\frac{1}{n}.
\end{equation}
Using Eq.~(\ref{611}) with the concentrated scattering function $\rho_n(x)$,
we arrive at
\begin{equation}
\label{1092}
f_{(S_n,X_n)}(s,x)=\Pr\left(T_n>\frac{s}{n}\right)\exp\left(-\frac{s}{n}\eta(
x)\right)\frac{\eta(x)\rho_n'(x)}{n}.
\end{equation}
With the definition of the concentrated scattering function $\rho_n(x)$,
this implies
\begin{equation}
f_{(S_n,X_n)}(s,x)=\Pr\left(T_n>\frac{s}{n}\right)\exp\left(-\frac{s}{n}\eta
(x)\right)\eta(x)\rho'(x).
\label{1093}
\end{equation}
Eqs.~(\ref{1091}) to (\ref{1093}) lead us to conclude that
\begin{equation}
\label{1094}
\lim_{n\rightarrow\infty}f_{(S_n,X_n)}(s,x)=\left[\lim_{n\rightarrow\infty}\Pr
\left(T_n>\frac{s}{n}\right)\right]\eta(x)\rho'(x),
\end{equation}%
with $s,x>0$. Now, using Eq.~(\ref{411}) with the concentrated scattering
function $\rho_n(x)$ we find
\begin{eqnarray}
\nonumber
&&\lim_{n\rightarrow\infty }\Pr\left(T_n>\frac{s}{n}\right)\\
&&=\lim_{n\rightarrow\infty}\exp\left(-\int_0^{\infty}\left[1-
\exp\left(-\eta(x)\frac{s}{n}\right)\right]\rho_n(dx)\right).
\hspace*{0.8cm}
\label{1095}
\end{eqnarray}
With the definition of the concentrated scattering function $\rho_n(x)$ this
leads to
\begin{eqnarray}
\nonumber
&&\lim_{n\rightarrow\infty }\Pr\left(T_n>\frac{s}{n}\right)\\
&&=\exp\left(-\int_0^{\infty}\left[\lim_{n\rightarrow\infty}
\frac{1-\exp\left(-\eta(x)t/n\right)}{1/n}\right]\rho(dx)\right).
\hspace*{0.8cm}
\label{1096}
\end{eqnarray}
Finally, with de l'Hospital's rule and Eq.~(\ref{600}) we obtain
\begin{eqnarray}
\nonumber
&&\lim_{n\rightarrow\infty }\Pr\left(T_n>\frac{s}{n}\right)\\
&&\hspace*{0.2cm}=\exp\left(-\int_0^{\infty}[\eta(x)t]\rho(dx)\right)=\exp(
-\lambda t).
\label{1097}
\end{eqnarray}
Combining together the limits of Eqs.~(\ref{1093}) and (\ref{1097})
proves Eq.~(\ref{601}).

\end{appendix}

\end{document}